\documentclass[aps,twocolumn,showpacs]{revtex4} 
\usepackage{latexsym}

\newcommand{\stac}[2]{\stackrel{\scriptscriptstyle {#1}}{#2}}

\begin{document}

\title{Gravity in Randall-Sundrum two D-brane model}

\author{Tetsuya Shiromizu$^{(1,2,3)}$, Keitaro Takahashi$^{(2)}$, Yoshiaki Himemoto$^{(2)}$ and 
Shuto Yamamoto$^{(1)}$ }

\affiliation{$^{(1)}$Department of Physics, Tokyo Institute of Technology, 
Tokyo 152-8551, Japan}

\affiliation{$^{(2)}$Department of Physics, The University of Tokyo,  Tokyo 
113-0033, Japan}

\affiliation{$^{(3)}$Advanced Research Institute for Science and Engineering, 
Waseda University, Tokyo 169-8555, Japan}

\date{\today}

\begin{abstract}
We analyse Randall-Sundrum two D-brane model by linear perturbation and then consider 
the linearised gravity on the D-brane. The qualitative contribution from the Kaluza-Klein 
modes of gauge fields to the coupling to the gravity on the brane will be addressed. 
As a consequence, the gauge fields localised on the brane are shown not to contribute to 
the gravity on the brane at large distances. Although the coupling between gauge fields 
and gravity appears in the next order, the ordinary coupling cannot be realised. 
\end{abstract}


\maketitle

\label{sec:intro}
\section{Introduction}

Recently the model construction of the inflation using D-brane has been initiated 
\cite{Dbrane} in braneworld context. See Refs. \cite{DBW1,DBW2,DBW3} for other related 
issues. However, the self-gravity of D-brane, which would be essential in considering 
D-brane cosmology, was not seriously considered there. On the other hand, 
Randall-Sundrum(RS) \cite{RSI,RSII} type model based on D-brane action has been considered 
in Refs. \cite{SKOT,SKT,OSKH,SHT}. In these papers, the bulk spacetime is described 
by type IIB supergravity compactified on $S^5$ \cite{ST} and the brane action is 
the Born-Infeld plus Chern-Simons action. Then the long wave approximation \cite{GE} 
was employed to discuss the low energy effective theory on the D-brane. Although the 
gauge field was assumed to be localised on the brane, the gravity on the D-brane was 
shown not to couple to it. 

In this paper we will reexamine the gravity on the branes by 
investigating the linear perturbation. So far, in the previous series
of paper on RS D-braneworld \cite{SKOT,SKT,OSKH,SHT}, 
the gradient expansion (long wave approximation) has been
employed to derive the effective theory on the brane. 
Furthermore, the form fields $B_2$ and $C_2$ are assumed to be closed, 
that is, $dB_2 = dC_2=0$, for simplicity. However, such assumption kills
the transverse tensor part of the form fields. In this paper, on the other hand, 
we will not impose such an assumption. Then we derive the linearised gravity on 
the D-branes. Our linear perturbation analysis 
follows Refs. \cite{GT,TM,Kim,harmonics}. As seen below, the contribution from the zero
mode of gauge fields does not have the usual form and is negligible at large distances.
This is consistent with the result from long wave approximation discussed in the
previous paper \cite{SHT} and the appendix of this paper.

The rest of this paper is organised as follows. In Sec. II, we describe the model which 
we consider here. In Sec. III, we formulate the ADM formalism for the current model. 
In Sec. IV, the linear perturbation analysis will be done and then derive the linearised 
gravitational equation on the brane. The contributions from massive modes of the form 
fields are also considered.  Finally we will give summary and discussion 
in Sec. V. In the appendix, for a comparison with the result obtained in linear perturbation 
analysis, we rederive the gravitational equation on 
the brane using the gradient expansion without assumption of $dB_2=dC_2=0$.

\section{Model}
\label{sec:model}

We consider the Randall-Sundrum model in type IIB supergravity compactified on 
$S^5$. The brane is described by Born-Infeld and Chern-Simons actions. So we begin with 
the following action 
%
\begin{eqnarray}
S & = & \frac{1}{2\kappa^2} \int d^5x {\sqrt {-{\cal G}}}\biggl[{}^{(5)}R-2\Lambda 
-\frac{1}{2}|H|^2 
\nonumber \\ 
& & -\frac{1}{2}(\nabla \chi)^2-\frac{1}{2}|\tilde F|^2-\frac{1}{2}|\tilde 
G|^2 \biggr] \nonumber \\ 
& & +S_{\rm brane}^{(+)}+S_{\rm CS}^{(+)}+S_{\rm brane}^{(-)}+S_{\rm CS}^{(-)} , 
\label{action} 
\end{eqnarray} 
%
where $H_{MNK}=\frac{1}{2}\partial_{[M}B_{NK]}$, 
$F_{MNK}=\frac{1}{2}\partial_{[M}C_{NK]}$, 
$G_{K_1 K_2 K_3 K_4 K_5}=\frac{1}{4!}\partial_{[K_1}D_{K_2 K_3 K_4 K_5]}$, 
$\tilde F = F + \chi H$ and $\tilde G=G+C \wedge H$. $M,N,K=0,1,2,3,4$. 
$B_{MN}$ and $C_{MN}$ are 2-form fields, and $D_{K_1 K_2 K_3 K_4}$ is
a 4-form field. $\chi$ is a scalar field. ${\cal G}_{MN}$ is the metric 
of five dimensional spacetime. 

$S_{\rm brane}^{(\pm)}$ is given by Born-Infeld action
%
\begin{eqnarray}
S_{\rm brane}^{(+)}=\gamma_{(+)} \int d^4x {\sqrt {-{\rm det}(h+{\cal F}^{(+)})}}, 
\end{eqnarray}
%
%
\begin{eqnarray}
S_{\rm brane}^{(-)}=\gamma_{(-)} \int d^4x {\sqrt {-{\rm det}(q+{\cal F}^{(-)})}}, 
\end{eqnarray} 
%
where $h_{\mu\nu}$ and $q_{\mu\nu}$ are the induced metric on the $D_{\pm}$-brane and 
%
\begin{eqnarray}
{\cal F}_{\mu\nu}^{(\pm)}=B_{\mu\nu}^{(\pm)}+(|\gamma_{(\pm)}|)^{-1/2}F_{\mu\nu}^{(\pm)}.
\end{eqnarray}
%
$F_{\mu\nu}$ is the $U(1)$ gauge field on the brane. Here $\mu,\nu=0,1,2,3$ and 
$\gamma_{(\pm)}$ are $D_{\pm}$-brane tension. 

$S_{\rm CS}^{(\pm)}$ is Chern-Simons action 
%
\begin{eqnarray}
S_{\rm CS}^{(+)} & = & \gamma_{(+)} \int d^4x {\sqrt {-h}} 
\epsilon^{\mu\nu\rho\sigma}\biggl[ \frac{1}{4}{\cal 
F}_{\mu\nu}^{(+)}C_{\rho\sigma}^{(+)}+\frac{\chi}{8}{\cal F}_{\mu\nu}^{(+)}{\cal F}_{\rho\sigma}^{(+)}
\nonumber \\
& & +\frac{1}{24}D_{\mu\nu\rho\sigma}^{(+)} \biggr], 
\end{eqnarray}
%
%
\begin{eqnarray}
S_{\rm CS}^{(-)} & = & \gamma_{(-)} \int d^4x {\sqrt {-q}} 
\epsilon^{\mu\nu\rho\sigma}\biggl[ \frac{1}{4}{\cal 
F}_{\mu\nu}^{(-)}C_{\rho\sigma}^{(-)}+\frac{\chi}{8}{\cal F}_{\mu\nu}^{(-)}{\cal 
F}_{\rho\sigma}^{(-)} 
\nonumber \\
& & +\frac{1}{24}D_{\mu\nu\rho\sigma}^{(-)} \biggr].
\end{eqnarray}
%
Here the brane charges are set equal to the brane tensions. Therefore, our model 
contains BPS state of D-branes.

\section{Basic equations}

In this section we write down the basic equations and boundary conditions. 
Let us perform (1+4)-decomposition 
%
\begin{eqnarray}
ds^2={\cal G}_{MN}dx^{M}dx^{N}=e^{2\phi (y,x)}dy^2+g_{\mu\nu}(y,x) dx^\mu dx^\nu,
\end{eqnarray}
%
where $y$ is the coordinate orthogonal to the brane. 
$D_+$-brane and $D_-$-brane are supposed to locate at $y=y^{(+)}=0$ and $y=y^{(-)}=y_0$. 
 
The spacelike ``evolutional" equations to the $y$-direction are 
%
\begin{eqnarray}
e^{-\phi} \partial_y K 
& = & {}^{(4)} R-\kappa^2 \biggl( {}^{(5)}T^\mu_\mu -\frac{4}{3}{}^{(5)}T^M_M \biggr) -K^2 
\nonumber \\
& & -e^{-\phi}D^2 e^\phi, 
\label{evoK}
\end{eqnarray}
%
%
\begin{eqnarray}
e^{-\phi} \partial_y \tilde K^\mu_\nu & = &  {}^{(4)}\tilde R^\mu_\nu 
-\kappa^2\biggl({}^{(5)}T^\mu_\nu 
-\frac{1}{4} 
\delta^\mu_\nu {}^{(5)}T^\alpha_\alpha \biggr)-K \tilde K^\mu_\nu \nonumber \\ 
& & ~~-e^{-\phi}[D^\mu D_\nu e^{\phi}]_{\rm traceless}, 
\label{traceless} 
\end{eqnarray}
%
%
\begin{eqnarray} 
\partial_y^2 \chi +D^2 \chi +e^\phi K\partial_y \chi-\frac{1}{2}H_{y\alpha\beta}\tilde 
F^{y\alpha\beta}=0, 
\end{eqnarray} 
%
%
\begin{eqnarray}
& & \partial_y X^{y\mu\nu}+e^\phi KX^{y\mu\nu}+D_\alpha \phi H^{\alpha\mu\nu}+D_\alpha H^{\alpha\mu\nu} 
\nonumber \\ 
& & ~~~~+\frac{1}{2}F_{y\alpha\beta}\tilde G^{y\alpha\beta\mu\nu}=0, 
\label{evoH} 
\end{eqnarray} 
%
%
\begin{eqnarray}
& & \partial_y \tilde F^{y\mu\nu}+e^\phi K \tilde F^{y\mu\nu}+D_\alpha \phi \tilde F^{\alpha\mu\nu}
+D_\alpha \tilde F^{\alpha\mu\nu}
\nonumber \\
& & ~~~~-\frac{1}{2}H_{y\alpha\beta}
\tilde G^{y\alpha\beta\mu\nu}=0,
\label{evoF}
\end{eqnarray}
%
%
\begin{eqnarray}
\partial_y \tilde G_{y \alpha_1 \alpha_2 \alpha_3 \alpha_4}
=e^\phi K\tilde  G_{y \alpha_1 \alpha_2 \alpha_3 \alpha_4},
\end{eqnarray}
%
where $X^{y\mu\nu}:=H^{y\mu\nu}+\chi \tilde F^{y\mu\nu}$ and the 
energy-momentum tensor is 
%
\begin{eqnarray}
&& \kappa^2\;{}^{(5)\!}T_{MN} =  \frac{1}{2}\biggl[ \nabla_M \chi \nabla_N \chi
-\frac{1}{2}g_{MN} (\nabla \chi)^2 \biggr]
\nonumber \\
& & ~~~~~~~~~~
+\frac{1}{4}\biggl[H_{MKL}H_N^{~KL}-g_{MN}|H|^2 \biggr] 
\nonumber \\
& & ~~~~~~~~~~
 +\frac{1}{4}\biggl[\tilde F_{MKL}\tilde
F_N^{~KL}-g_{MN}|\tilde F|^2
\biggr]
\nonumber \\
& & ~~~~~~~~~~
 +\frac{1}{96}\tilde G_{MK_1 K_2 K_3 K_4} \tilde G_{N}^{~~K_1
K_2 K_3 K_4}-\Lambda g_{MN}.
\nonumber \\
& & 
\end{eqnarray}
%
$K_{\mu\nu}$ is the extrinsic curvature, $K_{\mu\nu}=\frac{1}{2}e^{-\phi} \partial_y g_{\mu\nu}$. 
$\tilde K^\mu_\nu$ and ${}^{(4)}\tilde R^\mu_\nu$ are the traceless parts 
of $K^\mu_\nu$ and ${}^{(4)}R^\mu_\nu$, respectively. 
Here $D_\mu$ is the covariant derivative with respect to $g_{\mu\nu}$.

The constraints on $y={\rm const.}$ hypersurfaces are 
%
\begin{eqnarray}
& & -\frac{1}{2}\biggl[{}^{(4)}R-\frac{3}{4}K^2+\tilde K^\mu_\nu \tilde K^\nu_\mu \biggr]
=\kappa^2\:{}^{(5)\!}T_{yy}e^{-2\phi}, 
\label{conK}
\end{eqnarray}
%
%
\begin{eqnarray}
D_\nu K^\nu_\mu-D_\mu K = \kappa^2\:{}^{(5)\!}T_{\mu y}e^{-\phi},
\end{eqnarray}
%
%
\begin{eqnarray}
D_\alpha(e^{\phi} X^{y\alpha\mu})+\frac{1}{6}e^\phi F_{\alpha_1 \alpha_2 \alpha_3} 
\tilde G^{y \alpha_1 \alpha_2 \alpha_3 \mu}= 0, \label{con1}
\end{eqnarray}
%
%
\begin{eqnarray}
D_\alpha (e^{\phi} \tilde F^{y\alpha\mu})-\frac{1}{6}e^\phi H_{\alpha_1 \alpha_2 \alpha_3}
\tilde G^{y \alpha_1 \alpha_2 \alpha_3 \mu}=0, \label{con2}
\end{eqnarray}
%
%
\begin{eqnarray}
D^\alpha (e^{-\phi} \tilde G_{y \alpha \mu_1 \mu_2 \mu_3})=0.
\end{eqnarray}
%

Under $Z_2$-symmetry, the junction conditions at the brane located $y=y^{(\pm)}$ are 
%
\begin{eqnarray}
& & \Bigl[K_{\mu\nu} - g_{\mu \nu} K\Bigr]_{y=y^{(\pm)}}  =  \mp 
\frac{\kappa^2}{2} \gamma_{(\pm)} (g_{\mu\nu}-T^{(\pm)}_{\mu\nu} ) +O(T_{\mu\nu}^2) \label{omit} \\ 
& & H_{y\mu\nu}(y^{(\pm)},x)=\mp \kappa^2 \gamma_{(\pm)} e^\phi {\cal F}_{\mu\nu}^{(\pm)}, \\ 
& & \tilde F_{y\mu\nu}(y^{(\pm)},x)
=\mp \frac{\kappa^2}{2}\gamma_{(\pm)} e^\phi \epsilon_{\mu\nu\alpha\beta}{\cal F}^{(\pm)\alpha\beta}, \\ 
& & \tilde G_{y\mu\nu\alpha\beta}(y^{(\pm)},x) 
=\mp \kappa^2 \gamma_{(\pm)} e^\phi \epsilon_{\mu\nu\alpha\beta},\\ 
& & \partial_y \chi (y^{(\pm)},x) 
= \mp \frac{\kappa^2}{8}\gamma_{(\pm)} e^\phi \epsilon^{\mu\nu\alpha\beta}{\cal F}^{(\pm)}_{\mu\nu}{\cal 
F}_{\alpha\beta}^{(\pm)}. 
\end{eqnarray}
%
In the above 
%
\begin{eqnarray}
T^{(\pm)\mu}_{~~~~~\nu}={\cal F}^{(\pm)\mu\alpha}{\cal F}^{(\pm)}_{\nu \alpha} -\frac{1}{4}\delta^\mu_\nu 
{\cal F}_{\alpha\beta}^{(\pm)} {\cal F}^{(\pm) \alpha\beta}.
\end{eqnarray}
%

From the junction condition for $\chi$, we can omit the contribution of $\chi$ to the gravitational 
equation on the brane in the approximations which we will employ. Moreover, we omit the quadratic 
term in Eq. (\ref{omit}).

\section{Linearised gravity}

\subsection{Background}

The background bulk spacetime is five dimensional anti-deSitter spacetime
and its metric is given by 
%
\begin{eqnarray}
\stac{(0)}{g}_{\mu\nu}=a^2(y)\eta_{\mu\nu}=e^{-\frac{2y}{\ell}}\eta_{\mu\nu},
\end{eqnarray}
%
where 
%
\begin{eqnarray}
\frac{1}{\ell}=-\frac{1}{6}\kappa^2 \gamma_{(+)}=\frac{1}{6}\kappa^2 \gamma_{(-)}
:=-\frac{1}{6}\kappa^2 \gamma ,
\label{tension1}
\end{eqnarray}
%
and
%
\begin{eqnarray}
2\Lambda +\frac{5\kappa^4}{6}\gamma^2=0 \label{tune1}.
\end{eqnarray}
%
$\ell$ is the curvature radius of anti-deSitter  spacetimes. 
Eqs. (\ref{tension1}) and (\ref{tune1}) represent the Randall-Sundrum tuning 
and then the tension $\gamma_{(+)}$ and $\gamma_{(-)}$ have the same magnitude with 
opposite signature, $\gamma_{(+)}<0$ and $\gamma_{(-)}>0$. 

In addition,
%
\begin{eqnarray}
\tilde G_{y\alpha_1 \alpha_2 \alpha_3 \alpha_4}=-a^4 \kappa^2 
\gamma  \epsilon_{\alpha_1 \alpha_2 \alpha_3 \alpha_4},
\label{G}
\end{eqnarray}
%
where $\epsilon_{\alpha_1 \alpha_2 \alpha_3 \alpha_4}$ is the Levi-Civita 
tensor with respect to the induced metric $h_{\mu\nu}$ on the 
brane. Other form fields vanish in this order.

\subsection{Linear perturbation}

First we consider the linear perturbation for the evolutional equation of 
$B_{\mu\nu}$ and $C_{\mu\nu}$. In linear order Eqs. (\ref{evoH}) and (\ref{evoF}) become 
%
\begin{equation}
\partial_y  H_{y\mu\nu}
+ a^{-2}\partial_\alpha H_{~\mu\nu}^{\alpha} 
 +\frac{3}{\ell}  F_{y\alpha\beta} 
 \epsilon_{\mu \nu}^{~~\alpha\beta} =0,
\label{evo2}
\end{equation}
%
and
%
\begin{equation}
\partial_y  F_{y\mu\nu}
+ a^{-2}\partial_\alpha F_{~\mu\nu}^{\alpha} 
 -\frac{3}{\ell}  H_{y\alpha\beta} 
 \epsilon_{\mu \nu}^{~~\alpha\beta} =0,
\label{evo3}
\end{equation}
%
where $ F^\alpha_{~\mu\nu}=\eta^{\alpha\beta}F_{\beta\mu\nu}$. 

The constraint equations (\ref{con1}) and (\ref{con2}) are rewritten as 
%
\begin{eqnarray}
F_{\mu\nu\alpha}
= \frac{\ell}{6}\epsilon_{\mu \nu \alpha}^{~~~~~\beta} 
\partial^\rho H_{y \rho \beta},
\label{conF}
\end{eqnarray} 
%
and
%
\begin{eqnarray}
H_{\mu\nu\alpha}=- \frac{\ell}{6}\epsilon_{\mu \nu \alpha}^{~~~~~\beta} \partial^\rho F_{y \rho \beta}. \label{conH}
\end{eqnarray}
%

Here note that we can impose the following gauge conditions 
%
\begin{eqnarray}
B_{y\mu}=C_{y\mu}=0,
\end{eqnarray}
%
using the gauge transformations 
$B_{MN} \to B'_{MN}=B_{MN}+\partial_M \int^y_0 dy' B_{yN}(y',x)-\partial_N \int^y_0 dy' B_{yM}(y',x)$
and 
$C_{MN} \to C'_{MN}=C_{MN}+\partial_M \int^y_0 dy' C_{yN}(y',x)-\partial_N \int^y_0 dy' C_{yM}(y',x)$. 
Then
%
\begin{eqnarray}
H_{y\mu\nu} = \partial_y B_{\mu\nu}~~{\rm and}~~F_{y\mu\nu} = \partial_y C_{\mu\nu}.
\end{eqnarray}
%
$B_{\mu\nu}$ and ${}^{*}C_{\mu\nu}$ are decomposed to 
%
\begin{eqnarray}
B_{\mu\nu}=B_{\mu\nu}^{TT}+\partial_\mu B_\nu^{T}-\partial_\nu B_\mu^{T},
\end{eqnarray}
%
and
%
\begin{eqnarray}
{}^{*}C_{\mu\nu} & := & \frac{1}{2}\epsilon_{\mu\nu}^{~~~\alpha \beta}C_{\alpha \beta}
                        \nonumber \\
                 &  = & {}^{*}C_{\mu\nu}^{TT}+\partial_\mu {}^{*}C_\nu^{T}
                        -\partial_\nu {}^{*}C_\mu^{(T)},
\end{eqnarray}
%
where $\partial^\mu B_{\mu\nu}^{TT} = \partial^\mu {}^{*}C_{\mu\nu}^{TT}=
\partial^\mu B_\mu^T=\partial^\mu C_\mu^T=0  $. 

In momentum space, the field equations are 
%
\begin{eqnarray}
\partial_y^2 B_{\mu\nu}^{(m)TT}-a^{-2}k^2 B_{\mu\nu}^{(m)TT}+\frac{6}{\ell} 
\partial_y {}^{*}C_{\mu\nu}^{(m)TT}=0, \label{eq:BTT}
\end{eqnarray}
%
%
\begin{eqnarray}
\partial_y^2 {}^{*}C_{\mu\nu}^{(m)TT}+\frac{6}{\ell} \partial_y B_{\mu\nu}^{(m)TT}=0,
\label{eq:CTT}
\end{eqnarray}
%
%
\begin{eqnarray}
\partial_y^2 B_{\mu}^{(m)T}+\frac{6}{\ell} \partial_y {}^{*}C_{\mu}^{(m)T}=0,
\label{eq:BT}
\end{eqnarray}
%
and
%
\begin{eqnarray}
\partial_y^2 {}^{*}C_{\mu}^{(m)T}-a^{-2}k^2 {}^{*}C_{\mu}^{(m)T}+\frac{6}{\ell} 
\partial_y B_{\mu}^{(m)T}=0.
\end{eqnarray}
%
The constraint equations become 
%
\begin{eqnarray}
B_{\mu\nu}^{(m)TT}+\frac{\ell}{6} \partial_y {}^{*}C_{\mu\nu}^{(m)TT}=0, \label{eq:Bcon}
\end{eqnarray}
%
and
%
\begin{eqnarray}
{}^{*}C_{\mu}^{(m)T}+\frac{\ell}{6} \partial_y B_{\mu}^{(m)T}=0,\label{eq:Ccon}
\end{eqnarray}
%
which are consistent with Eqs. (\ref{eq:CTT}) and (\ref{eq:BT}). 

Using of Eq. (\ref{eq:Bcon}), Eq. (\ref{eq:BTT}) becomes 
%
\begin{eqnarray}
\partial_y^2 B_{\mu\nu}^{(m)TT}-a^{-2}k^2 B_{\mu\nu}^{(m)TT}-\frac{36}{\ell^2} 
B_{\mu\nu}^{(m)TT}=0. \label{eq:BTT2}
\end{eqnarray}
%
In the same way, Eq. (\ref{eq:CTT}) with Eq. (\ref{eq:Ccon}) give us 
%
\begin{eqnarray}
\partial_y^2 {}^{*}C_{\mu}^{(m)T}-a^{-2}k^2 {}^{*}C_{\mu}^{(m)T}-\frac{36}{\ell^2} 
{}^{*}C_{\mu}^{(m)T}=0. \label{eq:CTT2}
\end{eqnarray}
%

The junction conditions are 
%
\begin{eqnarray}
\partial_y B_{\mu\nu}^{TT}(y^{(\pm)},x)= -\partial_y {}^{*}C_{\mu\nu}^{TT}(y^{(\pm)},x)=
-\kappa^2 \gamma B_{\mu\nu}^{(\pm)TT},
\end{eqnarray}
%
and
%
\begin{eqnarray}
\partial_y B_{\mu}^{T}(y^{(\pm)},x)= -\partial_y {}^{*}C_{\mu}^{T}(y^{(\pm)},x)=
-\kappa^2 \gamma A_{\mu}^{(\pm)T}.
\end{eqnarray}
%
In the above ${\cal F}^{(\pm)}_{\mu\nu}$ is decomposed to 
%
\begin{eqnarray}
{\cal F}^{(\pm)}_{\mu\nu}(x) & = & 
B_{\mu\nu}^{(\pm)TT}+\partial_\mu B^{(\pm)T}_\nu - \partial_\nu B^{(\pm)T}_\mu
+ F_{\mu\nu} \nonumber \\
&  = & B_{\mu\nu}^{(\pm)TT}+\partial_\mu A^{(\pm)T}_\nu - \partial_\nu A^{(\pm)T}_\mu,
\end{eqnarray}
%
where $A_\mu ^T:= A_\mu +B_\mu^T $ and $F_{\mu\nu}= \partial_\mu A_\nu -\partial_\nu A_\mu$. 

Next we focus on the perturbed metric 
%
\begin{eqnarray}
ds^2 & = & (1+2\phi)dy^2+(\gamma_{\mu\nu}+h_{\mu\nu})dx^\mu dx^\nu \nonumber \\
     & = & (1+2\phi)dy^2+(a^2\eta_{\mu\nu}+h_{\mu\nu}^{TT}-\gamma_{\mu\nu} \psi )dx^\mu dx^\nu. 
\end{eqnarray}
%
where $h^{TT}_{\mu\nu}$ is transverse-traceless part of $h_{\mu\nu}$. 
Since the bulk background spacetime is anti-deSitter spacetime, the Green function is 
exactly the same with that of the Randall-Sundrum model. The difference is just
presence of the bulk form fields. Therefore we follow the argument of Ref. 
\cite{GT,TM} and then the perturbation $\bar h_{\mu\nu}$ in the Gaussian normal coordinate 
can be computed as 
%
\begin{eqnarray}
\bar h_{\mu\nu} = h_{\mu\nu}^{TT}- \gamma_{\mu\nu}
\Biggl(\psi -  \frac{2}{\ell}  \hat \xi^5 (x) \Biggr),
\end{eqnarray}
%
where $\hat \xi^5 (x)$ is a brane-bending mode (radion field) and 
\begin{widetext}
%
\begin{eqnarray}
h_{\mu\nu}^{TT}(y,x) & = & -2\kappa^2 \int d^4x'G_R(y,x;0,x')|\gamma| \Sigma_{\mu\nu}^{(+)}(x')
+2\kappa^2 \int d^4x'G_R(y,x;y_0,x')|\gamma| \Sigma_{\mu\nu}^{(-)}(x') \nonumber \\
& & -2\kappa^2 \int dy' d^4x' G_R(y,x;y',x')\delta {}^{(5)}T_{\mu\nu}(y',x'), 
\label{formalsol}
\end{eqnarray}
%
where 
%
\begin{eqnarray}
\Sigma_{\mu\nu}^{(\pm)}= T_{\mu\nu}^{(\pm)}
+\frac{1}{\kappa^2} \Biggl(\partial_\mu \partial_\nu \hat \xi^{5(\pm)}
-\frac{1}{4}\gamma_{\mu\nu} \partial^2 \hat \xi^{5(\pm)} \Biggr) .
\end{eqnarray}
%
The first and second terms in the right-hand side of Eq. (\ref{formalsol}) 
come from the $D_+$ and $D_-$ brane, respectively. 
The third one is the contribution from the bulk fields. 
In the above $\delta {}^{(5)} T_{\mu\nu}$ is the projected bulk stress tensor 
in the linear order. $G_R$ is the five dimensional retarded Green function 
%
\begin{eqnarray}
G_R (y,x;y',x')= G_R^{(0)} (y,x;y',x')+ G_R^{({\rm KK})} (y,x;y',x'),
\end{eqnarray}
%
where 
%
\begin{eqnarray}
G_R^{(0)} (y,x;y',x')=-\int \frac{d^4k}{(2\pi)^4} e^{ik \cdot (x-x')}
\frac{1}{\ell(1-a^2_0)} \frac{a(y)^2 a(y')^2}{{\bf k}^2-(\omega + i \epsilon)^2},
\end{eqnarray}
%
and
%
\begin{eqnarray}
G_R^{({\rm KK})} (y,x;y',x')=-\int \frac{d^4k}{(2\pi)^4} e^{ik \cdot (x-x')}
\int dm \frac{u_m(y) u_m(y')}{m^2+{\bf k}^2-(\omega + i \epsilon)^2}.
\end{eqnarray}
%
\end{widetext}
$G_R^{(0)}$ is the truncated retarded Green function for zero mode. $u_m(y)$ are 
the mode functions which are expressed by Bessel functions, 
$u_m (y) \propto J_1(m\ell)N_2(mz)-N_1(m\ell)J_2(mz)$. 

In the present model, the equation for $\hat \xi^5$ becomes 
%
\begin{eqnarray}
\partial^2 \hat \xi^5 (x) = \frac{\kappa^2}{6}T^{(+)}=0, \label{bend}
\end{eqnarray}
%
that is, the radion is a massless scalar field. 

The equation for $\psi$ comes from the Hamiltonian constraint equation:
%
\begin{eqnarray}
-\frac{3}{2}\frac{1}{a^2}\partial^2 \psi = \kappa^2 \delta {}^{(5)}T^y_y ,
\end{eqnarray}
%
where 
%
\begin{eqnarray}
\kappa^2 \delta {}^{(5)}T^y_y & = & \frac{1}{8}(H_{y \alpha \beta}H_y^{~\alpha \beta}
+\tilde F_{y \alpha \beta}\tilde F_y^{~\alpha\beta}) \nonumber \\
& & -\frac{1}{24}(H_{\mu\alpha\beta}H^{\mu\alpha\beta}
+\tilde F_{\mu \alpha \beta}\tilde F^{\mu \alpha \beta} ).
\end{eqnarray}
%

The relation between $\phi$ and $\psi$ comes from the traceless part of $(\mu,\nu)$-component 
of five dimensional Einstein equation 
%
\begin{eqnarray}
\psi - \phi \sim \kappa^2 (1/\partial^2)^2 \partial^\mu \partial^\nu [\delta {}^{(5)}T_{\mu\nu}]_{\rm traceless}.
\end{eqnarray}
%

\subsection{Zero-mode truncation}

In order to see the low energy effective (gravitational) theory on the D-brane, 
we will truncate zero-mode carefully. Let us focus on the zero mode for $B_{\mu\nu}$ 
and $C_{\mu\nu}$. Introducing new variables 
%
\begin{eqnarray}
\Psi_{\mu\nu}^{(\pm)}:=B_{\mu\nu}^{(0)} \pm {}^{*}C_{\mu\nu}^{(0)},
\end{eqnarray}
%
we obtain two linearly independent equations 
%
\begin{eqnarray}
\partial_y^2 \Psi^{(\pm)}_{\mu\nu} \pm \frac{6}{\ell}  \partial_y \Psi^{(\pm)}_{\mu\nu}=0.
\end{eqnarray}
%
The junction conditions are written by 
%
\begin{eqnarray}
\partial_y \Psi^{(+)}_{\mu\nu}(y^{(\pm)},x)=0,
\end{eqnarray}
%
%
\begin{eqnarray}
\partial_y \Psi^{(-)}_{\mu\nu}(y^{(\pm)},x)=-2\kappa^2 \gamma {\cal F}_{\mu\nu}^{(\pm)}.
\end{eqnarray}
%

First it is easy to see that the solutions to the equations for gauge fields are 
%
\begin{eqnarray}
\Psi^{(+)}_{\mu\nu}(y,x) = \alpha_{\mu\nu}(x),
\end{eqnarray}
%
and
%
\begin{eqnarray}
\Psi^{(-)}_{\mu\nu}(y,x) = \frac{2}{\ell}a^{-6}(y) {\cal F}^{(+)}_{\mu\nu}(x)+\beta_{\mu\nu}(x),
\end{eqnarray}
%
using the junction condition at $y=0$. $\alpha_{\mu\nu}(x)$ and $\beta_{\mu\nu}(x)$ are 
constant of integrations which is not arisen in $H_{y\mu\nu}$ and $\tilde F_{y\mu\nu}$:
%
\begin{eqnarray}
& & H_{y\mu\nu}^{(0)}(y,x)= \partial_y B_{\mu\nu}^{(0)}(y,x)= -\kappa^2 \gamma a^{-6}(y)
{\cal F}^{(+)}_{\mu\nu}, \\
& & \tilde F_{y\mu\nu}^{(0)}(y,x) = \partial_y C_{\mu\nu}^{(0)}(y,x) = 
-\frac{\kappa^2 \gamma}{2}a^{-6}(y)\epsilon_{\mu\nu}^{~~~\alpha\beta}
{\cal F}^{(+)}_{\alpha\beta}.
\end{eqnarray}
%

The remaining junction condition then implies 
the relation between gauge fields on the two branes 
%
\begin{eqnarray}
a_0^6 {\cal F}^{(-)}_{\mu\nu}={\cal F}^{(+)}_{\mu\nu},
\end{eqnarray}
%
and
%
\begin{eqnarray}
T_{0 \mu\nu}^{(-)}= a_0^{-14}T_{0\mu\nu}^{(+)}.\label{stressrelation} 
\end{eqnarray}
%

We also can compute the bulk stress tensor as 
\begin{widetext}
%
\begin{eqnarray}
\kappa^2 \delta {}^{(5)}T_{\mu\nu} & = & 
\frac{1}{2}a^{-2} \Biggl(H_{\mu y \alpha}H_{\nu}^{~~y \alpha}-\frac{1}{4}\eta_{\mu\nu}
H_{y \alpha \beta}H^{y \alpha \beta} \Biggr) 
+\frac{1}{2}a^{-2}\Biggl(\tilde F_{\mu y \alpha}\tilde F_{\nu}^{~~y \alpha}-\frac{1}{4}\eta_{\mu\nu}
\tilde F_{y \alpha \beta} \tilde F^{y \alpha \beta} \Biggr) \nonumber \\
& & +\frac{1}{4}a^{-4}\Biggl(  H_{\mu \alpha \beta}H_\nu^{~\alpha\beta}-\frac{1}{6}\eta_{\mu\nu} 
H_{\alpha \beta \rho} H^{\alpha \beta \rho}  \Biggr) 
+\frac{1}{4}a^{-4} \Biggl(  \tilde F_{\mu \alpha \beta}\tilde F_\nu^{~\alpha\beta}-\frac{1}{6}\eta_{\mu\nu} 
\tilde F_{\alpha \beta \rho} \tilde F^{\alpha \beta \rho}  \Biggr) \nonumber \\ 
& = & 
\biggl(\frac{6}{\ell} \biggr)^2 a^{-14}T_{\mu\nu}^{(+)}
+\frac{1}{4}a^{-4}\Biggl(  H_{\mu \alpha \beta}H_\nu^{~\alpha\beta}-\frac{1}{6}\eta_{\mu\nu} 
H_{\alpha \beta \rho} H^{\alpha \beta \rho}  \Biggr) 
+\frac{1}{4}a^{-4} \Biggl(  \tilde F_{\mu \alpha \beta}\tilde F_\nu^{~\alpha\beta}-\frac{1}{6}\eta_{\mu\nu} 
\tilde F_{\alpha \beta \rho} \tilde F^{\alpha \beta \rho}  \Biggr), \label{bulkstress}
\end{eqnarray}
%
where $H^{\alpha\beta\rho}= \eta^{\alpha\mu}\eta^{\beta\nu}\eta^{\rho\sigma}H_{\mu\nu\sigma}$. 
From Eq. (\ref{formalsol}) with Eqs. (\ref{bend}), (\ref{stressrelation}) and (\ref{bulkstress}) 
we finally obtain the following linearised equation on branes 
%
\begin{eqnarray}
\partial^2 \bar h_{\mu\nu}(y,x) & = &  \partial^2 h_{\mu\nu}^{TT} -\gamma_{\mu\nu} \partial^2 \psi 
\nonumber \\
 & = & - 2\kappa^2 \frac{a^2(y)}{\ell(1-a_0^2)} |\gamma|T^{(+)}_{\mu\nu}
+2\kappa^2 \frac{a^2(y)a^2_0}{\ell(1-a_0^2)} |\gamma|T^{(-)}_{\mu\nu}
-2 \frac{6}{\ell^2}\frac{a^2(y)(a_0^{-12}-1)}{1-a^2_0} T_{\mu\nu}^{(+)}  \nonumber \\
& & - 2 \frac{a_0^2}{\ell(1-a_0^2)} \int^{y_0}_0 dy \frac{1}{4}a^{-2}
\Biggl[H_{\mu \alpha \beta} H_\nu^{~\alpha \beta}+ 
\tilde F_{\mu \alpha \beta} \tilde F_\nu^{~\alpha \beta}  \Biggr]_{\rm traceless} 
+\frac{2}{3}\kappa^2 \eta_{\mu\nu}\Biggl( \delta {}^{(5)}T^y_y \Biggr)^{(0)}   \nonumber \\
 & = &- 2 \frac{a_0^2}{\ell(1-a_0^2)} \int^{y_0}_0 dy \frac{1}{4}a^{-2}
\Biggl[H_{\mu \alpha \beta} H_\nu^{~\alpha \beta}+ 
\tilde F_{\mu \alpha \beta} \tilde F_\nu^{~\alpha \beta}  \Biggr]_{\rm traceless} 
+\frac{2}{3}\kappa^2 \eta_{\mu\nu}\Biggl( \delta {}^{(5)}T^y_y \Biggr)^{(0)} \nonumber \\
& = & O(H_{\mu\nu\alpha}^2).
\end{eqnarray}
%
\end{widetext}
that is, $\partial^2 \bar h_{\mu\nu} (x) = O(H_{\mu\nu\alpha}^2)=O((\partial_\mu {\cal F}_{\nu\alpha})^2 ) $. 
The gauge fields localised on the branes does 
not appear as usual. The appropriate contribution from the boundary 
stress tensor is exactly cancelled out by that from the bulk stress tensor. 
In the above $(\cdots )^{(0)}$ represents the 
zero mode part. For example, $(\delta {}^{(5)}T^{y}_{y})^{(0)}$ is 
%
\begin{eqnarray}
(\delta {}^{(5)}T^y_y)^{(0)}= -\frac{1}{24}(H_{\mu\alpha\beta}H^{\mu\alpha\beta}
+\tilde F_{\mu \alpha \beta}\tilde F^{\mu \alpha \beta} ).
\end{eqnarray}
%

\subsection{Massive modes of form fields}

The mode functions for $B_{\mu\nu}^{(m)TT}$ etc. satisfying the junction condition at $y=y^{(+)}=0$ is 
%
\begin{eqnarray}
\psi_m = {\sqrt {m \ell e^{y/\ell}}} \frac{\alpha_m J_6(m \ell e^{y/ \ell})-\beta_m 
N_6(m \ell e^{y/\ell})}{{\sqrt {\alpha_m^2+\beta^2_m}}},
\end{eqnarray}
%
where 
%
\begin{eqnarray}
\alpha_m = m \ell N_5 (m \ell) -6 N_6(m \ell),
\end{eqnarray}
%
and
%
\begin{eqnarray}
\beta_m = m \ell J_5 (m \ell) -6 J_6(m \ell).
\end{eqnarray}
%
$m$ should be quantised by the junction condition at $y=y^{(-)}=y_0$. For $m\ell \gg1 $ and $m \ell 
e^{y^{(-)} /\ell} \gg 1$, we obtain the mass spectrum of 
%
\begin{eqnarray}
m^{(\pm)}_n \simeq  \frac{n \pi}{\ell (1-e^{y_0/\ell})}.
\end{eqnarray}
%
After determination of 
correct normalisation, for $m \ell \ll 1$, we can evaluate the contribution 
from massive modes to the right-hand side of 
Eq. (\ref{formalsol}) as 
%
\begin{eqnarray}
|\kappa^2 \delta {}^{(5)}T_{\mu\nu}| \simeq \frac{1}{\ell^2} \biggl( \frac{r_0}{r} \biggr)^3 \biggl( 
\frac{\ell}{r}\biggr)^{12} |T_{\mu\nu}^{(+)}|  \ll \frac{1}{\ell^2}|T_{\mu\nu}^{(+)}|,
\end{eqnarray}
%
where $r_0$ is spatial scale of support of form field. 
Thus even if we consider the contribution from
the massive modes, they 
will be negligible at low energy scale.

\section{Summary and discussion}
\label{sec:summary}

In this paper we derived the linearised gravitational equation on the D-brane
and then it turned out that the gauge fields do not couple to the gravity
on the brane at zero modes in a usual way. Instead, an unusual couplings appear and
it is negligible at large distances. We also discussed the contribution 
from the Kaluza-Klein modes which was shown to be also negligible at large distances. 

The model which we considered is minimum extension of Randall-Sundrum type model to the 
supergravity-like one. Therein $Z_2$ symmetry and RS tuning are assumed. In this model, RS tuning 
corresponds to the condition of equality of brane tension and charge. It is likely 
that D-brane in BPS state does not provide us the realistic model for the braneworld. 
As analysed in Ref. \cite{SKT}, on the other hand, it was shown that 
the coupling of the gravity to the gauge fields will appears and the coupling constant is 
proportional to the cosmological constant. Therefore non-BPS state will be important for 
braneworld cosmology. We can also consider the cases without $Z_2$-symmetry. For example, a 
model considered in Ref. \cite{Dbrane} does not have $Z_2$-symmetry. So the careful study 
of the self-gravitational effect for such a mode will be important. 


\section*{Acknowledgements}

TS thank Y. Imamura for useful discussion. 
The work of TS was supported by Grant-in-Aid for Scientific
Research from Ministry of Education, Science, Sports and Culture of 
Japan(No.13135208, No.14740155 and No.14102004). 
The works of YH and KT were supported by JSPS.

\appendix

\section{Long-wave approximation}
\label{sec:approximation}

In this appendix, we approximately solve the bulk field equations by long wave 
approximation (gradient expansion \cite{GE}) and derive the effective theory on the 
brane. The equation obtained here will 
include the non-linear effect. Thus we can obtain the same result with one obtained in 
Sec. IV if we linearise the equation. This appendix can be regarded as the extension of 
previous work \cite{SHT} into the general cases where we do not impose $dC_2=dB_2=0$. 

In the case with bulk fields we must
carefully use the geometrical projection method \cite{SMS} because the projected Weyl
tensor $E_{\mu\nu}$ contains the leading effect from the bulk fields. 

The bulk metric is written again as,
%
\begin{eqnarray}
ds^2=e^{2\phi (x)}dy^2+g_{\mu\nu}(y,x) dx^\mu dx^\nu. 
\end{eqnarray}
%
The induced metric on the brane will be denoted by 
$h_{\mu\nu}:=g_{\mu\nu}(0,x)$ 
and then
%
\begin{eqnarray}
g_{\mu\nu}(y,x)
=a^2(y,x)\Bigl[h_{\mu\nu}(x)+\stac{(1)}{g}_{\mu\nu}(y,x)+\cdots\Bigr].
\end{eqnarray}
%
In the above $\stac{(1)}{g}_{\mu\nu}(0,x)=0 $ and $a(0,x)=1$. In a similar way, 
the extrinsic curvature is expanded as 
%
\begin{eqnarray}
K^\mu_\nu = \stac{(0)}{K^\mu_\nu}+ \stac{(1)}{K^\mu_\nu}+\stac{(2)}{K^\mu_\nu}+ 
\cdots.
\end{eqnarray}
%
The small parameter is $\epsilon = (\ell /L)^2 \ll 1$,
where $L$ and $\ell$ are the curvature scale on the brane 
and the bulk anti-deSitter curvature scale, respectively. 
%
%

It is easy to obtain the zeroth order solutions. Without derivation 
they are given by the Randall-Sundrum set up;
Eqs.(\ref{tension1}) and (\ref{tune1}). Then
%
\begin{eqnarray}
\stac{(0)}{K^\mu_\nu}=-\frac{1}{\ell}\delta^\mu_\nu,
\end{eqnarray}
%
%
\begin{eqnarray}
\stac{(0)}{g}_{\mu\nu}=a^2(y,x)h_{\mu\nu}(x)=e^{-\frac{2d(y,x)}{\ell}}h_{\mu\nu}(x),
\end{eqnarray}
%
where 
%
\begin{eqnarray}
d(y,x) = \int^y_0 dy e^{\phi(x)}. 
\end{eqnarray}
$\tilde G_{y\alpha_1 \alpha_2 \alpha_3 \alpha_4}$ is also given by Eq.(\ref{G}). 
%
%


Next we consider the first order equations. 
The first order equations for $\tilde F_{y\mu\nu}$ and $H_{y\mu\nu}$ are 
%
\begin{eqnarray}
\partial_y \stac{(1)}{\tilde F}_{y\mu\nu}-\frac{1}{2a^4} 
\stac{(1)}{H}_{y\alpha\beta} \tilde G_{y\rho\sigma\mu\nu}h^{\alpha\rho}h^{\beta\sigma}=0, 
\end{eqnarray}
%
and
%
\begin{eqnarray}
\partial_y \stac{(1)}{H}_{y\mu\nu}+\frac{1}{2a^4}\stac{(1)}{\tilde F}_{y\alpha\beta}\tilde 
G_{y\rho\sigma\mu\nu}h^{\alpha\rho}h^{\beta\sigma}=0.
\end{eqnarray}
%
Together with the junction conditions on $D_+$-brane the solutions are given by 
%
\begin{eqnarray}
\stac{(1)}{H}_{y\mu\nu}(y,x)=-\kappa^2 \gamma a^{-6}e^\phi {\cal F}_{\mu\nu}^{(+)},
\end{eqnarray}
%
and
%
\begin{eqnarray}
\stac{(1)}{\tilde F}_{y\mu\nu}(y,x)=-\frac{\kappa^2}{2}\gamma a^{-6} e^\phi 
\epsilon_{\mu\nu\rho\sigma}{\cal F}_{\alpha\beta}^{(+)}h^{\rho\alpha}h^{\sigma\beta}.
\end{eqnarray}
%
The remaining junction conditions on $D_-$-brane imply the relation between ${\cal F}^{(+)}_{\mu\nu}$ 
and ${\cal F}^{(-)}_{\mu\nu}$ as 
%
\begin{eqnarray}
{\cal F}_{\mu\nu}^{(-)}=a_0^{-6}{\cal F}_{\mu\nu}^{(+)},
\end{eqnarray}
%
and then 
%
\begin{eqnarray}
T_{\mu\nu}^{(-)}= a_0^{-14} T_{\mu\nu}^{(+)}, \label{emtensor}
\end{eqnarray}
%
where $a_0=a(y_0,x)=e^{-d_0(x)/\ell}$ and $d_0(x):=d(y_0,x)$. 

Let us first 
substitute the junction conditions for $H_{y\mu\nu}$ and $\tilde F_{y\mu\nu}$ on the $D_+$ brane 
into the constraint equations of Eqs. (\ref{con1}) and (\ref{con2}). Then we see 
%
\begin{eqnarray}
& & {\cal D}^\mu \Biggl( {\cal F}_{\mu\nu}^{(+)}-\frac{1}{2}\epsilon_{\mu\nu\alpha\beta}C^{\alpha\beta} \Biggr)=0, \label{max1} \\
& & \epsilon^{\mu\nu\alpha\beta} {\cal D}_\nu ({\cal F}^{(+)}_{\alpha\beta}-B_{\alpha\beta})=0, \label{max2}
\end{eqnarray}
%
where ${\cal D}_\mu$ is the covariant derivative with respect to $h_{\mu\nu}$.

Using these results the evolutional equation for the traceless part 
of the extrinsic curvature is 
%
\begin{eqnarray}
e^{-\phi} \partial_y \stac{(1)}{\tilde K^\mu_\nu} & = & 
-\stac{(0)}{K}\stac{(1)}{\tilde K^\mu_\nu}+
\tilde R^\mu_\nu (g) -\kappa^4 \gamma^2 a^{-16} T^{(+)\mu}_\nu \nonumber \\
& & ~~-e^{-\phi}[ D^\mu D_\nu e^\phi]_{\rm traceless}, 
\end{eqnarray}
%
where 
%
\begin{eqnarray}
\tilde R^\mu_\nu (g) = \frac{1}{a^2}\Biggl[ 
R^\mu_\nu (h) +\frac{2}{\ell} {\cal D}^\mu {\cal D}_\nu d +\frac{2}{\ell^2}{\cal D}^\mu d {\cal D}_\nu d 
\Biggl]_{\rm traceless},
\end{eqnarray}
%
and
%
\begin{eqnarray}
D^\mu D_\nu e^\phi & = &  \frac{1}{a^2} {\cal D}^\mu {\cal D}_\nu e^\phi 
+\frac{1}{a^2 \ell} \Biggl({\cal D}^\mu e^\phi {\cal D}_\nu d \nonumber \\
& &  +{\cal D}^\mu d {\cal D}_\nu e^\phi 
-\delta^\mu_\nu {\cal D}^\alpha d {\cal D}_\alpha e^\phi  \Biggr).
\end{eqnarray}
%
$R^\mu_\nu (h)=h^{\mu\alpha}R_{\alpha\nu}(h)$ 
is the Ricci tensor with respect to $h_{\mu\nu}$ 
and $T^\mu_\nu= h^{\mu\alpha}T_{\alpha\nu}$. 

The solution is summarised as 
%
\begin{eqnarray}
\stac{(1)}{\tilde K^\mu_\nu}(y,x)  & =  &  -\frac{\ell}{2a^2}{}^{(4)}\tilde R^\mu_\nu (h) 
+\frac{1}{2}\kappa^2 \gamma  a^{-16}T^{(+)\mu}_{~~~\nu} \nonumber \\
& & -a^{-2}\Biggl[{\cal D}^\mu {\cal D}_\nu  d - \frac{1}{\ell} {\cal D}^\mu d {\cal D}_\nu d 
\Biggr]_{\rm traceless} \nonumber \\
& & + \frac{\chi^\mu_\nu (x)}{a^4}, 
\label{sol}
\end{eqnarray}
%
where $\chi^\mu_\nu$ is the ``integration of constant". 

The solution to the trace part of the extrinsic curvature is 
%
\begin{eqnarray}
\stac{(1)}{K}(y,x) & = & -\frac{\ell}{6a^2} {}^{(4)}R(h) \nonumber \\
 & & -\frac{1}{a^2}{\cal D}^2 d +\frac{1}{a^2 \ell} ({\cal D}d)^2. \label{trace}
\end{eqnarray}
%

On the $D_+$-brane Eqs (\ref{sol}) and (\ref{trace}) becomes 
%
\begin{eqnarray}
{}^{(4)}\tilde R^\mu_\nu (h) = \frac{2}{\ell} \chi^\mu_\nu (x) 
\end{eqnarray}
%
and
%
\begin{eqnarray}
0=\stac{(1)}{K}(0,x)=-\frac{\ell}{6} {}^{(4)}R(h)
\end{eqnarray}
%
They correspond to the Einstein equation on the brane obtained in Ref. \cite{SMS} and 
$ \chi^\mu_\nu$ is projected Weyl tensor $E_{\mu\nu}$. For the moment, 
$\chi^\mu_\nu (x)$ is unknown term. 

On $D_- $-brane, Eq. (\ref{sol}) becomes 
%
\begin{eqnarray}
\frac{\kappa^2}{2}\gamma T^{(-) \mu}_{~~~\nu} & = &  -\frac{\ell}{2a_0^2} {}^{(4)}\tilde R^\mu_\nu (h) 
+\frac{\kappa^2}{2}a_0^{-16}\gamma T^{(+)\mu}_{~~~\nu} \nonumber \\
& &-\frac{1}{a_0^2} \Biggl[ {\cal D}^\mu {\cal D}_\nu d_0-\frac{1}{\ell}{\cal D}^\mu d_0 
{\cal D}_\nu d_0 \Biggr]_{\rm traceless} \nonumber \\
& & +\frac{\chi^\mu_\nu (x)}{a_0^4}.
\end{eqnarray}
%
and
%
\begin{eqnarray}
0=\stac{(1)}{K}(y_0,x)= -\frac{1}{a_0^2}{\cal D}^2 d_0+\frac{1}{a_0^2 \ell}({\cal D}d_0)^2.
\end{eqnarray}
%

All together we obtain the Einstein equation on $D_+$ brane 
%
\begin{eqnarray}
(a_0^{-2}-1)G_{\mu\nu} (h) = \frac{2}{\ell} \Biggl[{\cal D}_\mu {\cal D}_\nu d_0 
-\frac{1}{\ell} {\cal D}_\mu d_0 {\cal D}_\nu d_0  \Biggr]_{\rm traceless}.
\end{eqnarray}
%
The equation for radion becomes 
%
\begin{eqnarray}
{\cal D}^2 d_0-\frac{1}{\ell} ({\cal D}d_0)^2=0.
\end{eqnarray}
%
Defining 
%
\begin{eqnarray}
\Psi = 1-e^{-2d_0/\ell}~~{\rm and}~~ \omega(\Psi) = \frac{3}{2} \frac{\Psi}{\Psi-1},
\end{eqnarray}
%
we rewrite down the Einstein equation as 
%
\begin{eqnarray}
G_{\mu\nu}(h) = \Biggl[ \frac{1}{\Psi} {\cal D}_\mu {\cal D}_\nu \Psi + \frac{\omega }{\Psi^2}
{\cal D}_\mu \Psi {\cal D}_\nu \Psi \Biggr]_{\rm traceless}.
\end{eqnarray}
%
and
%
\begin{eqnarray}
{\cal D}^2 \Psi + \frac{1}{2\omega +3 } \frac{d \omega}{d \Psi} ({\cal D}\Psi)^2=0.
\end{eqnarray}
%
Thus the contribution from the gauge fields to gravity on the brane does not exist at 
low energy scale.


\end{document}